\documentstyle[12pt,a4]{article}
\begin{document}
\begin{center}
{\bf MULTISCALE ANALYSIS OF A DAVYDOV MODEL WITH AN HARMONIC LONG RANGE 
INTERACTION OF KAC-BAKER TYPE}
\vskip .3cm
{\it Dan Grecu, Anca Vi\c sinescu}\\
{\it Department of Theoretical Physics}\\
{\it National Institute for Physics and Nuclear Engineering}\\
{\it P.O.Box MG-6, M\v agurele, Bucharest, Romania}\\
{\it e-mail: dgrecu@theor1.theory.nipne.ro}\\
{\it ~~~~~~~~~avisin@theor1.theory.nipne.ro} 
\end{center}

\vskip 1cm
\begin{abstract}
 The classical equation of motion of a Davydov model in a coherent 
state approximation is analyzed using the multiple scales method. An
exponentially decaying long range interaction (Kac-Baker model) was
included. In the first order, the dominant amplitude has to be a solution 
of the nonlinear Schr\"odinger equation (NLS). In the next order the 
second amplitude
satisfies an inhomogeneous linearized NLS equation, the inhomogeneous term
depending only on the dominant amplitude. In order to eliminate possible
secular behaviour the dominant amplitude has to satisfy also the next
equation in the NLS hierarchy (a complex modified KdV equation).
When the second order derivative of the dispersion relation
vanishes the scaling of the slow space variable has to be changed, and a
generalized NLS equation with a third order derivative is found for the
dominant amplitude. As the coefficient of the third derivative is small a
perturbational approach is used to discuss the equation.
A complete solution is given when the dominant amplitude is the 
one-soliton solution of the NLS equation.
\end{abstract}
\vskip 1cm
{\bf 1. Introduction}
\vskip .5cm
Many quasi-one-dimensional molecular systems are very complicated 
structures built from complexes of atoms - we call them "molecules" - 
connected by hydrogen bonds. As an example we mention the complicate 
structure of $\alpha$-helix in protein. A very simple model consists in 
replacing the three coupled chains of the complex $\alpha$-helix 
structure by only one chain of the form $\cdot\cdot\cdot 
H-N-C=O-H-N-C=O\cdot\cdot\cdot$. Also only one of the intramolecular 
excitations - that corresponding to the amide I oscillation - is taken 
into account (further on they will be called vibrons). An acoustic 
phonon field describing the oscillations of the molecules along the 
chain is also introduced. As the amide I oscillation energy depends on 
the stretching of the adjacent hydrogen bond an anharmonic interaction 
between vibrons and phonons appears.

It is now easy to write an 
Hamiltonian of Fr\"ohlich type for this very simplified model. For a 
system with a conserved number of vibrons the model was proposed by 
Davydov \cite{1}-\cite{4} more than 25 years ago. Eliminating this 
constraint of fixed number of vibrons the model was extended by Takeno 
\cite{5},\cite{6},\cite{3} at the beginning of the eighties.

The Hamiltonian of Davydov's model is given by
\begin{eqnarray}\label{1}
&\hat H&=\sum_{n}E B^{+}_{n}B_{n}-\sum_{n\ne 
m}J_{mn}(B^{+}_{n}B_{m}+B^{+}_{m}B_{n}) \nonumber \\
&+&\sum_{n}({1\over 2m}\hat 
p^{2}_{n}+{1\over 2}w(\hat u_{n+1}-\hat u_{n})^{2})+\chi \sum_{n}(\hat 
u_{n+1}-\hat u_{n})B^{+}_{n}B_{n}.
\end{eqnarray}
Here $B_{n},B^{+}_{n}$ are the annihilation/creation operators for 
vibronic excitation of energy $E$ in the cell $n$, $J_{mn}$ term takes 
into account the long range interaction between vibrons, $\hat u_{n}, 
\hat p_{n}$ are the displacement operator and the corresponding 
conjugate momentum of the $n$-th molecule, $m$ is an effective mass of 
the molecule, $w$ the elastic constant and $\chi$ a coupling constant
describing the nonlinear 
interaction between vibrons and phonons. As an example of the long 
range interaction between vibrons we shall consider an exponentially 
decreasing model (Kac-Baker model) \cite{7}
\begin{equation}\label{2}
J_{mn}=J{1-r\over 2r}e^{-\gamma\vert m-n\vert}, ~~~~r=e^{-\gamma}.
\end{equation}
Since the pioneering paper of Sarker and Krumhansl \cite{sar} this 
model was intensively used by a series of authors to investigate the 
thermodynamic properties and the soliton characteristics in several 
nonlinear 1-D systems \cite{rem} - \cite{gai}.

It is a general belief that a coherent state approximation is 
very suitable for describing extended localized states in such systems. 
Such an hypothesis was done by Davydov. His ansatz for the state 
vector is
\begin{equation}\label{3}
\vert\Psi(t)>=\sum_{n}a_{n}(t)B^{+}_{n} exp\left( -{i\over 
\hbar}\sum_{j}\beta_{j}(t)\hat p_{j}-\pi_{j}(t)\hat u_{j}\right)\vert 
0>,
\end{equation}
$\vert 0>$ being the vacuum state both for vibrons and phonons. Here 
$a_{n}(t), \beta_{n}(t), \pi_{n}(t)$ are now $c$-numbers, time 
depending, which will be determined from a variational principle. 
Using the average value of $\hat H$ as 
Hamiltonian in the classical equations of motion one gets
\begin{eqnarray}\label{4}
i\hbar \dot{a}_{n}&=&Ea_{n}-\sum^{\infty}_{p=1}J_{p}(a_{n+p}+a_{n-p})+\chi 
(\beta_{n+1}-\beta_{n})a_{n} \nonumber \\
M\ddot{\beta}&=&w(\beta_{n+1}-2\beta_{n}+\beta_{n-1})+\chi(\vert 
a_{n}\vert^{2}-\vert a_{n-1}\vert^{2}).
\end{eqnarray}
In an adiabatic approximation $(\dot\beta_{n}\rightarrow 0)$ from the 
second eq. (\ref{4}) we get
$$
\beta_{n+1}-\beta_{n}=-{\chi\over w}\vert a_{n}\vert^{2},
$$
which introduced into the first gives
\begin{equation}\label{5}
i\hbar \dot{a_{n}}=Ea_{n}-\sum J_{p}(a_{n+p}+a_{n-p})-{\chi^{2}\over 
w}\vert a_{n}\vert^{2}a_{n}.
\end{equation}
a typical self-trapping equation.
\vskip 1cm
 
{\bf 2. Multiple scale approach}
\vskip .5cm

The linearized problem admits plane wave solutions with the dispersion 
relation
\begin{equation}\label{6}
\hbar \omega (k)=E-J{1-r\over 2r}\sum^{\infty}_{p=1}e^{-\gamma p}\cos 
klp=E-J{1-r\over 2r}\left({\sinh \gamma\over cosh \gamma -\cos 
kl}-1\right).
\end{equation}
A plane wave solution with constant amplitude exists even for the 
nonlinear equation (5), but with an amplitude depending dispersion 
relation  (Stokes waves)
\begin{equation}\label{7}
\hbar \omega (k)=E-J{1-r\over 2r}\sum_{p}e^{-\gamma p}\cos klp 
-{\chi^{2}\over w}\vert a\vert^{2}.
\end{equation}
It is well known that these solutions are unstable at small modulation 
of the amplitude (Benjamin Feir instability) \cite{8}. To see this in 
our case we write
$$a_{n}=a(1-{\cal A}_{n}(t))e^{i(kln-\omega t)}$$
where ${\cal A}_{n}(t)$ are satisfying the following system of linear 
equations
$$i\hbar \dot{\cal A}_{n}=2\sum J_{p}{\cal A}_{n} \cos klp -\sum 
J_{p}({\cal A}_{n+p}e^{iklp}+{\cal A}_{n-p}e^{-iklp})-{\chi^{2}\over w}\vert 
a\vert^{2}({\cal A}_{n}+{\cal A}^{\star}_{n}).$$
Looking for plane wave solutions
$${\cal A}_{n}=c e^{i(\nu ln-\Omega t)}+d e^{-i(\nu ln -\Omega t)}$$
the instability, $\Omega^{2}<0$, appears when
$$\left(\sum_{p}J_{p}\cos klp (1-\cos \nu 
lp)\right)\left(\sum_{p}J_{p}\cos klp(1-\cos\nu lp)-{\chi^{2}\over 
w}\vert a\vert^{2}\right)\le 0$$
which is always satisfied at small wave numbers $\nu$.

In order to describe this process of amplitude modulation, the multiple 
scales approach (reductive perturbative method) \cite{9}-\cite{cam} is used. 
The solution is written as an asymptotic expansion in a small parameter 
$\epsilon$.
\begin{equation}\label{12}
a_{n}=e^{i(kln-\omega t)}\sum_{j}\epsilon^{j}A_{j}(\xi,t_{2},t_{3},...)
\end{equation}
where the amplitudes $A_{j}$ depend on some slow variables 
$\xi,t_{2},t_{3},... $ In order to see how these can be defined let us 
expand the dispersion relation around a point $k_{0} ~(k=k_{0}+\Delta 
k)$
$$
\omega (k)=\omega_{0}+\omega_{1} \Delta k+\omega_{2}\Delta k^{2}+ ...,
~~~~~~~~~~~\omega_{n}={1\over n! }{d^{n}\omega(k)\over dk^{n}}\vert 
_{k=k_{0}}$$
and considers $\Delta k=\mu \epsilon^{\alpha}$. Then the phase of the 
plane wave can be written
\begin{equation}\label{13}
kln-\omega t=(k_{0}ln-\omega_{0}t)+\mu 
\epsilon^{\alpha}(ln-\omega_{1}t)-\mu^{2}\omega_{2}\epsilon^{2\alpha}t-
\mu^{3}\omega_{3}\epsilon^{3\alpha}t+...
\end{equation}
Two distinct situations can appear. In the first case $\omega_{2}\ne 0 
$ (we shall call it "a normal situation") and the second when 
$\omega_{2}=0$. This last situation is referred as the "zero dispersion 
point" case (ZDP). In the normal situation, with $\alpha =1$, and the 
following definition of the slow variables 
\begin{equation}\label{14}
\xi = \epsilon (ln-v_{g}t), ~~~~~~~~~\tau_{2}=\epsilon^{2}t, ~~~~~~~~~
\tau_{3}=\epsilon^{3} t, ...
\end{equation}
it is easily seen that a competition between nonlinearity,  
${\partial A_{1}\over \partial \tau_{2}}$ and ${\partial^{2} A_{1}\over 
\partial \tau_{2}^{2}}  $ will occur in order $\epsilon^{3}$. 
Introducing these in (\ref{5}) we ask that the equation has to be satisfied 
in each order of $\epsilon$. In the first order in $\epsilon$ the 
equation is linear and we obtain the dispersion relation (\ref{6}). In the 
next order $\epsilon^{2}$ the equation is satisfied if the velocity 
$v_{g}$ is given by $v_{g}=\omega_{1}$. In the order $\epsilon^{3}$ we 
found that the leading amplitude $A_{1}$ has to satisfy the well known 
nonlinear Schr\"odinger equation (NLS) 
\begin{equation}\label{15}
{\partial A_{1}\over \partial \tau_{2}}+\omega_{2}\left({\partial^{2} 
A_{1}\over \partial \xi^{2}}+2c\vert A_{1}\vert^{2}A_{1}\right)=0, 
~~~~c={q\over 2\omega_{2}}, ~~~~q={\chi^{2}\over \hbar w}.
\end{equation}
Recently \cite{10} the analysis for Takeno's model was 
extended to the next order $\epsilon^{4}$. Similarly, in the present 
case, a nonhomogeneous linear 
equation satisfied by the next amplitude $A_{2}$ is obtained, namely
\begin{equation}\label{16}
i{\partial A_{2}\over \partial \tau_{2}}+\omega_{2}{\partial^{2}A_{2}\over 
\partial \xi^{2}}+q(A_{1}^{2}A_{2}^{\star}+2\vert 
A_{1}\vert^{2}A_{2})=-i{\partial A_{1}\over \partial 
\tau_{3}}+i\omega_{3}{\partial^{3}A_{1}\over \partial \xi^{3}}.
\end{equation}
In the left hand side (lhs) we recognize the linearized NLS equation 
(l-NLS eq.), while the nonhomogeneity in the right hand side (rhs) 
contains only the dominant amplitude $A_{1}$, solution of the NLS eq. 
In solving the equation (\ref{16}) we are confronted with two distinct 
problems. Firstly we have to take care to eliminate any secular 
behaviours raised by the presence of the nonhomogeneity. They can appear 
from terms in the rhs which are members of the null space of the l-NLS eq. 
As is well known the symmetries of the NLS eq. are solutions of the l-NLS
eq., so the dangerous terms in the rhs of (\ref{16}) are to be found between 
them. Such a symmetry is easily identified in (\ref{16}) namely
\begin{equation}\label{17}
\sigma_{3}=-\left({\partial^{3}A_{1}\over \partial \xi^{3}}+6c\vert 
A_{1}\vert^{2}{\partial A_{1}\over \partial \xi}\right)
\end{equation}
and the possible secular behaviour is eliminated if the $\tau_{3}$ 
dependence of $A_{1}$ is given by
\begin{equation}\label{18}
-{\partial A_{1}\over \partial \tau_{3}}
+\omega_{3}\left({\partial^{3} A_{1} \over 
\partial \xi^{3}}+6c\vert A_{1}\vert^{2}{\partial A_{1} \over \partial 
\xi}\right)=0.
\end{equation}
This is a complex modified KdV equation and is the second equation in 
the hierarchy associated to NLS eq. As all the equations in the 
hierarchy have the same spectral problem the $\tau_{3}$ time dependence 
of $A_{1}$ can appear only in the initial positions and phases 
characterizing the solution $A_{1}$ of the NLS eq. After this 
"renormalization" of the dominant amplitude $A_{1}$ we remain with a 
nonhomogeneous linear equation, free of secularities, for the second 
amplitude $A_{2}$
\begin{equation}\label{19}
i{\partial A_{2}\over 
\partial\tau_{2}}+\omega_{2}{\partial^{2}A_{2}\over 
\partial\xi^{2}}+q(A^{2}_{1}A^{\star}_{2}+2\vert 
A_{1}\vert^{2}A_{2})=-6i\omega_{3}c\vert A_{1}\vert^{2}{\partial 
A_{1}\over \partial\xi}.
\end{equation}
A simple method to solve it when $A_{1}$ is the one-soliton solution of 
NLS eq. will be given in the next section. More comments will be 
presented there.

In the last years several papers have discussed the significance of the 
higher order approximations to a nonlinear dynamical problem, which in 
the lowest relevant order is described by a completely integrable 
equation \cite{11}-\cite{13}, and the role played by the next equations 
from the corresponding hierarchies in order to eliminate possible secular 
behaviours. Our results are in completely agreement with these 
discussions.
\vskip 1cm
{\bf 3. ZDP case}
\vskip .5cm

Let us discuss now the case of zero dispersion point.
For models in which the carrier wave is of the form of 
a lattice plane wave such a point $k_{0}$ exist always in the first 
Brillouin zone. It corresponds to the maximum group velocity of the 
localized excitation. For the dispersion relation (\ref{6}) it is given by 
\begin{equation}\label{20}
\cos k_{0}l={1\over 2}\left(\sqrt{\cosh^{2}\gamma +8}-\cosh 
\gamma\right).
\end{equation}
The analysis presented in the previous paragraph has to be slightly 
modified. In the phase (\ref{13}) the $\omega_{2}$ term is missing. Then the 
next term has to be used to define the time variable $\tau_{2}$ and 
consequently we have to consider $\alpha={2\over 3}$. The slow 
variables are now given by
\begin{equation}\label{21}
\xi=\epsilon^{{2\over 3}}(ln-\omega_{1}t), 
~~~~~~\tau_{2}=\epsilon^{2}t.
\end{equation}
In order $\epsilon^{3}$ we shall have a competition between the 
nonlinearity, the time derivative ${\partial A_{1}\over \partial 
\tau_{2}}$ and the third order spatial derivative 
${\partial^{3}A_{1}\over \partial \xi^{3}}$.
Also the second derivative ${\partial^{2} A_{1}\over \partial \xi^{2}}$
can contribute in the same order $\epsilon^{3}$ because it is 
multiplied by $\omega_{2}(k)$, which expanded around $k_{0}$ brings an 
extra $\epsilon^{{2\over 3}}$ dependence $(\omega_{2}(k)=3\omega_{3}\mu 
\epsilon^{{2\over 3}}+...)$. Collecting all these contributions of 
$\epsilon^{3}$ order we arrive at the following equation satisfied by 
the dominant amplitude $A_{1}$
\begin{equation}\label{22}
i{\partial A_{1}\over \partial 
\tau_{2}}+3\mu\omega_{3}{\partial^{2}A_{1}\over \partial 
\xi^{2}}+q\vert A_{1}\vert^{2}A_{1}=i\omega_{3}{\partial^{3}A_{1}\over 
\partial\xi^{3}}.
\end{equation}
Introducing dimensionless quantities
\begin{equation}\label{23}
X={1\over\sqrt{6\mu}}k_{0}\xi, ~~~~~~T=\omega_{3}k^{3}_{0}\tau_{2}, 
~~~~~~\Psi=\sqrt{{\chi^{2}\over \hbar\omega_{3}k^{3}_{0}w}}A_{1}
\end{equation}
the equation (\ref{22}) becomes
\begin{equation}\label{24}
I{\partial \Psi\over \partial T}+{1\over 2}{\partial^{2}\Psi\over 
\partial X^{2}}+\vert \Psi\vert^{2}\Psi=i\beta{\partial^{3}\Psi\over 
\partial X^{3}}
\end{equation}
where $\beta=(6\mu)^{-{3\over 2}}$. This is a NLS eq. perturbed with a 
third order derivative term. As $\beta$ is small the eq. 
(\ref{24}) will be solved by a perturbation approach. We consider
\begin{equation}\label{25}
\Psi\rightarrow \Psi+\beta \delta \Psi
\end{equation}
where now $\Psi $ is a solution of the unperturbed NLS eq. Also we 
introduce a slow time variable $\tau=\beta T$. 
In first order in $\beta$ we get
\begin{equation}\label{26}
i{\partial \delta\Psi\over \partial T}+{1\over 
2}{\partial^{2}\delta\Psi\over \partial X^{2}}+(2\vert 
\Psi\vert^{2}\delta\Psi+\Psi^{2}\delta\Psi^{\star})=-i{\partial\Psi\over
 \partial\tau}+i{\partial^{3}\Psi\over \partial X^{3}}.
\end{equation}
This equation is  exactly of the same form as (\ref{16}). The possible 
secular behaviour is eliminated requiring that the $\tau$ dependence of 
$\Psi$ is given by the complex mKdV equation
\begin{equation}\label{27}
-{\partial \Psi\over \partial \tau}+{\partial^{3}\Psi\over \partial 
X^{3}}+6\vert\Psi\vert^{2}{\partial\Psi\over \partial X}=0.
\end{equation}
We remain with a linear nonhomogeneous equation for $\delta \Psi$ 
namely
\begin{equation}\label{28}
i{\partial \delta\Psi\over \partial T}+{1\over 
2}{\partial^{2}\delta\Psi\over \partial X^{2}}+(2\vert 
\Psi\vert^{2}\delta\Psi+\Psi^{2}\delta\Psi^{\star})=-6i\vert\Psi\vert^{2
}{\partial\Psi\over\partial X}
\end{equation}
which is free of secularities.

Further we shall give a complete solution for the case when $\Psi$ is 
the one-soliton solution of the NLS eq. If $z=u+iv$ is the eigenvalue 
of the spectral problem the one soliton solution is given by
\begin{equation}\label{29}
\Psi=2v{e^{-i\Phi}\over \cosh \Theta}, 
~~~~~~\Phi=2uX+2(u^{2}-v^{2})T+\Phi_{0}, ~~~~~~\Theta=2v(X-X_{0}+2uT).
\end{equation}
As mentioned before, the $\tau$-dependence can appear only in the 
initial phase $\Phi_{0}$ and the initial position $X_{0}$. Introducing 
(\ref{29}) in (\ref{27}) it is easily find that
\begin{equation}\label{30}
{d\Phi_{0}\over d\tau}=8u(u^{2}-3v^{2}), ~~~~~~{dX_{0}\over 
d\tau}=4(3u^{2}-v^{2})
\end{equation}
which lead to a linear dependence of $\Phi_{0}$ and $X_{0}$ on the slow 
time variable $\tau$.

As is well known the NLS eq. is invariant under Galilei transformation. 
The same is true also for the l-NLS eq., so without any lost of 
generality we can solve (\ref{28}) in the reference frame where the soliton 
is at rest $(u=0)$ and then through the Galilei transformation
\begin{equation}\label{31}
X=X^{\prime}+2uT^{\prime}, ~~~~~~~~T=T^{\prime}, 
~~~~~~~~\delta\Psi=\delta\Psi^{\prime}e^{(2iuX^{\prime}+2iu^{2}T^{\prime
})}
\end{equation}
we can find the solution in the laboratory system. The one-soliton 
solution in its own reference frame is given by
\begin{equation}\label{32}
\Psi=2v{e^{2iv^{2}T}\over \cosh 2vX}.
\end{equation}
We are looking for solutions of the form
\begin{equation}\label{33}
\delta\Psi=ie^{2iv^{2}T}Y(X)
\end{equation}
with $Y(X)$ a real function. With $2vX\rightarrow X$ and introducing 
the new variable $\rho =\tanh X$ the equation satisfied by $Y$
becomes
\begin{equation}\label{35}
{d\over d\rho}\left((1-\rho^{2}){dY\over d\rho}\right)+(2-{1\over 
1-\rho^{2}})Y=48v^{2}\rho\sqrt{1-\rho^{2}}.
\end{equation}
In the lhs we recognize the equation for the associated Legendre 
polynomials. The two linear independent solutions are $P^{1}_{1}$ and 
$Q_{1}^{1}$
\begin{equation}\label{36}
P^{1}_{1}=-\sqrt{1-\rho^{2}}, 
~~~~~~Q^{1}_{1}=-\sqrt{1-\rho^{2}}\left({1\over 2}ln{1+\rho\over 
1-\rho}+{\rho\over 1-\rho^{2}}\right).
\end{equation}
We write the solution of the nonhomogeneous equation (\ref{35}) as a linear 
superposition
\begin{equation}\label{37}
Y(\rho )=\alpha (\rho) P^{1}_{1}(\rho)+\beta (\rho)Q^{1}_{1}(\rho),
\end{equation}
where $\alpha (\rho), \beta (\rho)$ are coefficients to be determined.
A general way to find them is to consider
\begin{equation}\label{38}
{d\alpha\over d\rho}=fQ_{1}^{1}, ~~~~~~~~~~~~~{d\beta\over 
d\rho}=fP^{1}_{1}.
\end{equation}
Then using the Wronskian expression
it is easily found that 
\begin{equation}\label{40}
f(\rho )=-24 v^{2}\rho\sqrt{1-\rho^{2}}.
\end{equation}
Then the equations for $\alpha$ and $\beta$ are easily integrated 
giving
\begin{equation}\label{42}
\alpha (\rho)=-3v^{2}(1-\rho^{2})^{2} ln{1+\rho \over 
1-\rho}+6v^{2}\rho (1+\rho^{2}), ~~~~\beta (\rho)=6v^{2}(1-\rho^{2}).
\end{equation}
In integrating the equation for $\beta$ an integration constant was 
determined from the condition $\beta (\pm 1)=0$. This ensure us to have 
$Y(\rho )$ finite everywhere for $\rho \in [-1,+1].$ Now introducing 
(\ref{42}) in (\ref{37}) a very simple form for $Y(\rho )$ is found, namely
\begin{equation}\label{43}
Y(\rho )=-12 v^{2}\rho\sqrt{1-\rho^{2}}.
\end{equation}
As $Y(\rho )$ is zero both in the origin and at $\rho =\pm 1 
~~~(X\rightarrow \pm \infty )$ it has a maximum at 
$\rho=\pm{1\over\sqrt{2}}$, which transformed in the $X$ variable gives 
$X={1\over 2v}ln(1+\sqrt{2}).$ This maximum is similar with that found 
in numerical simulations and other theoretical treatments of the 
similar ZDP problem of pulse propagation in nonlinear optical fibers 
\cite{14}-\cite{19}.

\vskip 1cm
{\bf 4. Conclusions}
\vskip .5cm

Solitonic type excitations in a Davydov model are investigated. 
The multiple scales method is used to study the space-time 
modulation of the amplitude. As expected the dominant amplitude $A_{1}$ 
is satisfying a NLS eq. In the next order the second amplitude $A_{2}$ 
is given by the solution of a nonhomogeneous linearized NLS equation. 
Possible secular behaviours are generated if symmetries of NLS eq. are 
identified in the rhs of this equation. They are eliminated if the NLS 
solution satisfies also the next eq. in the NLS hierarchy (a complex 
mKdV equation). The ZDP case is also investigated using the same 
method of multiple scales, but with another definition of the slow 
spatial variable. The dominant amplitude satisfies a modified NLS 
equation containing a third order derivative term. As its coefficient 
is a small quantity a perturbational approach is used. The case of the 
one-soliton solution of the NLS eq. is fully solved. Possible secular 
behavior is eliminated if the one-soliton solution satisfies also the 
next equation in the NLS hierarchy. A linear $\tau_{3}$ dependence of 
the initial phase and position of the soliton is determined. The 
remaining equation, by a suitable transformation, is reduced to the 
equation for associated Legendre polynomials. The complete solution is 
found if the NLS soliton is at rest. By a Galilei transformation the 
solution for moving soliton is obtained. More complicated situations 
are under investigation.

\vskip .5cm

{\it {\bf Acknowledgments} Helpful discussions with Dr. A.S. C\^arstea 
are kindly acknowledged. This research was supported under the Grant 
No. 6008/2000 with the National Agency for Science, Technology and 
Innovation (ANSTI).}

\end{document}